\documentclass[11pt]{article}
\usepackage[margin=0.90in]{geometry}
\usepackage{amsmath,amssymb,amsthm,mathtools}
\usepackage{microtype}
\usepackage{booktabs,array}
\usepackage{enumitem}
\usepackage{tikz}
\usetikzlibrary{arrows.meta,positioning,calc,fit,backgrounds}
\usepackage{xcolor}
\definecolor{linkblue}{RGB}{24,76,130}
\definecolor{figblue}{RGB}{45,96,150}
\definecolor{figgray}{RGB}{95,105,115}
\usepackage[colorlinks=true,linkcolor=linkblue,citecolor=linkblue,urlcolor=linkblue]{hyperref}
\usepackage[nameinlink,capitalize]{cleveref}
\usepackage{mathpazo}
\usepackage[T1]{fontenc}
\usepackage{authblk}
\usepackage{caption}
\setlist{nosep,leftmargin=*}
\newtheorem{theorem}{Theorem}[section]
\newtheorem{lemma}[theorem]{Lemma}

\newtheorem{corollary}[theorem]{Corollary}
\newtheorem{definition}[theorem]{Definition}
\newtheorem{remark}[theorem]{Remark}

\newcommand{\C}{\mathcal C}
\newcommand{\G}{\mathcal G}
\newcommand{\Pau}{\mathcal P}

\newcommand{\Q}{\mathbb Q}
\newcommand{\Z}{\mathbb Z}

\newcommand{\Tr}{\operatorname{Tr}}

\newcommand{\mdepth}{\operatorname{mdepth}}
\newcommand{\FieldDepth}{\operatorname{Fdepth}}
\newcommand{\AlphabetDepth}{\operatorname{Adepth}}
\newcommand{\ResourceDepth}{\operatorname{Rdepth}}
\newcommand{\Tcost}{\mathcal T}

\newcommand{\Compile}{\operatorname{Compile}}

\newcommand{\vecop}{\operatorname{vec}}

\title{\bfseries Arithmetic Depth, Minimal Gate Alphabets, \\and Exact Program-State Resources\\
for Single-Qubit Dyadic Programs}
\author[1]{Tony Newton}
\author[2]{Dan-Adrian German}
\affil[1]{Newton Astro Labs, London, United Kingdom\\\texttt{tony.newton79@gmail.com}}
\affil[2]{Luddy SICE, Indiana University, Bloomington, IN, USA\\\texttt{dgerman@iu.edu}}
\date{October 2026}

\begin{document}
\maketitle

\begin{abstract}
We develop an exact resource semantics for a dyadic fragment of single-qubit quantum programming that originates in normalized mist-state rewriting.  A primitive dyadic source state of reduced coordinate $m/2^D$ has minimum normalized-midpoint depth $D$ and is Clifford-equivalent to the standard phase-program state for the rotation $U_z(m\pi/2^D)$.  We then move from this axial fragment to the projective single-qubit Clifford-cyclotomic tower
\[
  \widehat\G_D\subset SO(3),\qquad
  \G_D=\langle \C,U_z(\pi/2^D)\rangle,\qquad
  K_D=\Q\!\left(2\cos\frac{\pi}{2^D}\right).
\]
Our main arithmetic result is the relative intersection theorem
\[
  \widehat{\G}_D\cap SO(3,K_d)=\widehat{\G}_d,
  \qquad 1\le d\le D.
\]
Its proof combines the denominator-exponent canonical form of Forest--Gosset--Kliuchnikov--McKinnon with a parity lemma for the quadratic real-cyclotomic tower.  Independently, we prove that finite adaptive stabilizer processing cannot enlarge the field generated by resource-state Pauli expectations.  Consequently, for every non-Clifford channel represented by a unitary in the dyadic Clifford-cyclotomic union, three a priori different quantities coincide: the smallest real cyclotomic field containing its Bloch matrix, the shallowest exact dyadic gate alphabet up to global phase, and the shallowest dyadic program-state layer sufficient for exact stabilizer-assisted implementation.  At that minimal layer, the known fixed-alphabet synthesis theorem gives minimum primitive-rotation count.  The resulting pair ``arithmetic depth, optimal count'' is therefore an exact, lexicographically optimal static resource effect for this single-qubit programming model.
\end{abstract}

\section{Introduction}

Fault-tolerant quantum programming uses two descriptions of non-Clifford computation that are usually studied separately.  In one description a program is a circuit over a gate alphabet, such as Clifford+$T$ or a finer Clifford-cyclotomic alphabet.  In the other, non-Clifford gates are supplied by consumable quantum states and measurement-based injection.  The first viewpoint leads naturally to exact synthesis; the second to magic-state resource theories.  A programming language that exposes both views should ideally assign a resource effect that is simultaneously a static property of source syntax, a lower bound on physical resources, and an exact predictor of the smallest gate alphabet required by the compiled program.

\paragraph{Projective convention.}
All compilation and resource claims in this paper concern quantum channels, so global phase is physically irrelevant.  Accordingly, gate-alphabet depth is defined through the Bloch image in $SO(3)$ (equivalently in $PU(2)$), not through literal membership of a particular phase representative in $U(2)$.  This convention is essential in the relative-descent argument below and matches the phase-insensitive cost convention of exact single-qubit synthesis.

The starting point here is the mist-state rewriting formalism developed in \cite{German2026,NewtonGerman2026}.  Its certified version separates raw amplitude collection from normalization, gives a canonical amplitude normal form, and proves that nested equal-weight normalized mists generate dyadic geodesic barycenters \cite{NewtonGerman2026}.  In particular, the nested state at angle $\pi/8$ is the positive Hadamard eigenstate.  Rather than treating this state as an isolated fixed point, we use the dyadic-geodesic theorem as a source language for program states.

The paper then leaves the pedagogical setting.  The target is the exact single-qubit dyadic Clifford-cyclotomic union.  Forest, Gosset, Kliuchnikov and McKinnon proved a unique canonical form and an optimal exact synthesis algorithm for every \emph{fixed} Clifford-cyclotomic alphabet \cite{Forest2015}.  Their denominator-exponent theorem contains enough arithmetic information to answer a different question: given a unitary known to lie somewhere in the nested dyadic tower, what is the \emph{smallest} alphabet in which it can live?  We prove that this minimum is determined exactly by the smallest real cyclotomic field containing the Bloch matrix.

The second ingredient is operational.  Beverland, Campbell, Howard and Kliuchnikov established powerful non-Clifford lower bounds using stabilizer nullity and a dyadic monotone, including adaptive stabilizer computations \cite{Beverland2020}.  We use a complementary invariant: not a scalar amount of magic, but the number field generated by Pauli expectation values.  Stabilizer processing is arithmetically closed over that field.  This produces an exact field-reachability obstruction that is insensitive to how many shallower resource states are supplied.

\paragraph{Main result.}
For $D\ge1$ let
\[
 \G_D:=\left\langle \C,U_z\!\left(\frac{\pi}{2^D}\right)\right\rangle,
 \qquad
 K_D:=\Q\!\left(2\cos\frac{\pi}{2^D}\right),
\]
with $K_1=\Q$ and with unitaries identified up to global phase through their Bloch representations.  For a non-Clifford channel $\mathcal U(\rho)=U\rho U^\dagger$ with $\widehat U\in\bigcup_D\widehat\G_D$, define
\begin{align*}
 \FieldDepth(U)&:=\min\{D:\widehat U\in SO(3,K_D)\},\\
 \AlphabetDepth(U)&:=\min\{D:\widehat U\in\widehat\G_D\},
\end{align*}
and let $\ResourceDepth(U)$ be the minimum possible maximum depth of dyadic program states in any finite adaptive stabilizer protocol implementing $\mathcal U$ exactly.  We prove
\begin{equation}\label{eq:headline}
 \boxed{\FieldDepth(U)=\AlphabetDepth(U)=\ResourceDepth(U).}
\end{equation}
This equality remains meaningful for programs outside every finite level of the Clifford hierarchy: gate-hierarchy membership is not compositional above the Clifford group, whereas the arithmetic resource depth remains finite and exact.

At the common minimum $D_*$, the canonical synthesis of \cite{Forest2015} minimizes the number $\Tcost_{2^{D_*}}(U)$ of primitive $\pi/2^{D_*}$ rotations.  Hence
\[
 \operatorname{QEff}(U):=\bigl(D_*,\Tcost_{2^{D_*}}(U)\bigr)
\]
is a lexicographically optimal resource signature: first it minimizes the deepest dyadic resource layer, then it minimizes primitive-rotation count within that layer.

\begin{figure}[t]
\centering
\begin{tikzpicture}[node distance=8mm and 10mm, every node/.style={font=\small},
 box/.style={draw=black,rounded corners=2pt,fill=white,minimum height=8mm,align=center,inner sep=4pt},
 arr/.style={-{Latex[length=2.4mm]},thick,draw=figblue}]
\node[box] (mist) {normalized mist term\\$m/2^D$};
\node[box,right=of mist, xshift=2mm] (state) {program state\\$q_{m,D}$};
\node[box,right=of state, xshift=4mm] (inj) {adaptive\\gate injection};
\node[box,right=of inj, xshift=3mm] (gate) {dyadic channel\\$\widehat U\in\widehat\G_D$};
\node[box,below=of gate] (field) {Bloch field\\$K_D$};
\node[box,left=of field, xshift=-7mm] (lower) {stabilizer\\field obstruction};
\node[box,left=of lower, xshift=-2mm] (type) {exact effect\\$(D_*,\Tcost_*)$};
\draw[arr] (mist)-- node[above,font=\scriptsize]{depth} (state);
\draw[arr] (state)-- node[above,font=\scriptsize]{Clifford} (inj);
\draw[arr] (inj)-- node[above,font=\scriptsize]{exact} (gate);
\draw[arr] (gate)-- node[right,font=\scriptsize]{field} (field);
\draw[arr] (field)-- node[below,font=\scriptsize]{lower bound} (lower);
\draw[arr] (lower)-- node[below,font=\scriptsize]{equality} (type);
\draw[arr] (type.west) -| ([xshift=-4mm]mist.west) -- (mist.west);
\end{tikzpicture}
\caption{The proof architecture.  The source language supplies dyadic program states; arithmetic closure gives an operational lower bound; relative cyclotomic descent identifies that lower bound with the smallest exact gate alphabet.}
\label{fig:architecture}
\end{figure}
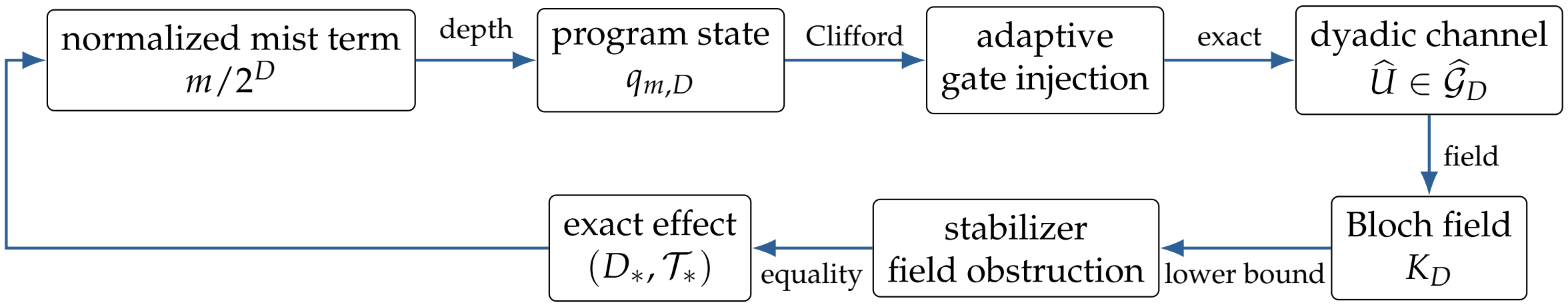

\paragraph{Relation to prior work.}
Magic-state injection and small-angle resource states are established tools \cite{CampbellOGorman2016,ZhengZangKubica2026}.  Complete one-qubit Clifford-hierarchy structure is now known \cite{deSilvaLautsch2025}, and dyadic quantum fragments have complete rewriting systems in other formalisms \cite{Vilmart2024}.  Quantum type systems already reason about Clifford structure, $T$ gates and selected injection circuits \cite{Sundaram2022}, while resource-aware languages infer quantitative upper bounds such as circuit width \cite{Colledan2025}.  The contribution here is narrower: an exact equality between arithmetic field depth, minimum dyadic gate-alphabet depth and minimum dyadic program-state depth, together with an exact static resource effect that remains meaningful beyond the finite Clifford hierarchy.  Earlier rotation-group structure goes back to Radin--Sadun, and later arithmetic-group work studies the ambient Clifford-cyclotomic groups; recent multiqubit synthesis results show that ancillas can change ambient ring-synthesis questions.  Our theorem is instead a relative single-qubit descent statement inside the nested dyadic tower.

\section{Dyadic mist states as program states}\label{sec:mist}

We first isolate the fragment of mist-state semantics needed later.  Let
\[
 q(x):=\cos\!\left(\frac{\pi x}{2}\right)|0\rangle
      +\sin\!\left(\frac{\pi x}{2}\right)|1\rangle,
 \qquad 0\le x\le1.
\]
For equal-norm real states on the shorter arc, normalized addition is the geodesic midpoint \cite{NewtonGerman2026}:
\[
 \mu(q(x),q(y)):=N(q(x)+q(y))=q\!\left(\frac{x+y}{2}\right).
\]

\begin{definition}[Normalized-midpoint depth]
A normalized-midpoint term is generated from endpoint leaves $0,1$ and the binary constructor $\mu$.  Its denotation is the dyadic coordinate obtained from $0,1$ by repeated averaging.  The minimum depth of any term denoting $x$ is denoted $\mdepth(x)$.
\end{definition}

\begin{theorem}[Exact dyadic depth]\label{thm:dyadicdepth}
Let $x=m/2^D\in(0,1)$ be in lowest dyadic form, so $m$ is odd.  Then
\[
 \mdepth(x)=D.
\]
\end{theorem}

\begin{proof}
By induction, every midpoint tree of maximum depth $d$ evaluates to a rational whose reduced denominator divides $2^d$.  Hence $m/2^D$ cannot occur below depth $D$.  Conversely, use a complete binary tree with $2^D$ leaves, label exactly $m$ leaves by $1$ and the remainder by $0$, and average level by level.  The root evaluates to $m/2^D$.
\end{proof}

The relevant source state is therefore
\begin{equation}\label{eq:qmd}
 q_{m,D}:=q(m/2^D)
 =\cos\frac{m\pi}{2^{D+1}}|0\rangle
 +\sin\frac{m\pi}{2^{D+1}}|1\rangle.
\end{equation}
For $m=1,D=2$ this is the positive Hadamard eigenstate highlighted in \cite{German2026,NewtonGerman2026}.

\subsection{Uniform state--gate conversion}
Define the standard phase-program state
\[
 |A_\theta\rangle:=\frac{|0\rangle+e^{i\theta}|1\rangle}{\sqrt2},
 \qquad U_z(\theta):=\operatorname{diag}(1,e^{i\theta}).
\]
A single fixed Clifford relates the two encodings.

\begin{lemma}[Uniform Clifford equivalence]\label{lem:SH}
For every real $\theta$,
\[
 SH|A_\theta\rangle
 =e^{i\theta/2}
 \left(\cos\frac\theta2|0\rangle+\sin\frac\theta2|1\rangle\right).
\]
In particular, with $\theta=m\pi/2^D$, $q_{m,D}$ is Clifford-equivalent to $|A_\theta\rangle$.
\end{lemma}

\begin{proof}
Expand $H|A_\theta\rangle$ and use
$1+e^{i\theta}=2e^{i\theta/2}\cos(\theta/2)$ and
$1-e^{i\theta}=-2ie^{i\theta/2}\sin(\theta/2)$.  The phase gate $S=\operatorname{diag}(1,i)$ removes the factor $-i$ from the second component.
\end{proof}

The standard one-bit injection circuit then makes the program-state interpretation operational.  For $|\psi\rangle=a|0\rangle+b|1\rangle$, use the data qubit as control of a CNOT whose target is $|A_\theta\rangle$, then measure the ancilla in the computational basis.  The two unnormalized data branches are
\begin{align}
 0 &: \quad U_z(\theta)|\psi\rangle/\sqrt2,\label{eq:inj0}\\
 1 &: \quad e^{i\theta}U_z(-\theta)|\psi\rangle/\sqrt2.\label{eq:inj1}
\end{align}
Thus the second branch is corrected by $U_z(2\theta)$.  This doubled-angle feed-forward is the familiar small-angle magic-state cascade \cite{CampbellOGorman2016}.

For the primitive dyadic phase $x=0.b_1\cdots b_D$, doubling removes the first binary digit:
\[
 2x=b_1+0.b_2\cdots b_D.
\]
Since $U_z(\pi)=Z$ is Clifford,
\begin{equation}\label{eq:shift}
 U_z(\pi x)^2
 =Z^{b_1}U_z\!\left(\pi\,0.b_2\cdots b_D\right)
 \quad\text{up to global phase}.
\end{equation}
Hence a failed injection consumes one bit of the canonical dyadic source word while emitting only a free Pauli correction.  The worst-case number of non-Clifford injection stages is $D-1$, whereas the expected number is
\[
 1+\frac12+\cdots+\frac1{2^{D-2}}
 =2-2^{-(D-2)}<2.
\]
We use this axial compiler later only for the upper-bound direction of the resource theorem; the main arithmetic result applies to arbitrary multiaxis words in the dyadic Clifford-cyclotomic group.

\section{Arithmetic closure of stabilizer protocols}\label{sec:closure}

We next give the operational lower-bound mechanism.  The invariant is field-valued rather than a scalar magic monotone.

For an $n$-qubit state $\rho$ and Pauli $P\in\Pau_n$, write
\[
 r_P(\rho):=\Tr(P\rho).
\]
For a finite resource family $\mathcal R=\{\rho_1,\ldots,\rho_s\}$ define
\begin{align}
 A(\mathcal R)&:=\Z\!\left[\frac12,\{r_P(\rho_j)\}_{j,P}\right],\label{eq:ring}\\
 K(\mathcal R)&:=\operatorname{Frac}A(\mathcal R).\label{eq:fracfield}
\end{align}

\begin{theorem}[Stabilizer arithmetic closure]\label{thm:closure}
Consider any finite adaptive protocol generated by Clifford unitaries, stabilizer ancillas, Pauli measurements, deterministic classical feed-forward and discarding, with nonfree input resources $\mathcal R$.  No external operation carrying additional real or algebraic parameters is supplied for free.  For every measurement history $s$, every Pauli coordinate of the unnormalized branch $\widetilde\rho_s$ lies in $A(\mathcal R)$.  If $p_s=\Tr\widetilde\rho_s\neq0$, then every Pauli coordinate of the normalized branch $\rho_s=\widetilde\rho_s/p_s$ lies in $K(\mathcal R)$.
\end{theorem}

\begin{proof}
The claim is preserved by each primitive operation.  Clifford conjugation sends a Pauli to $\pm$ another Pauli and therefore only permutes and signs coordinates.  Stabilizer ancillas have Pauli expectations in $\{0,\pm1\}$.  Tensor products multiply Pauli coordinates.  If $M$ is the measured Pauli and $\Pi_b=(I+(-1)^bM)/2$, then for any Pauli $Q$,
\[
 \Pi_bQ\Pi_b=
 \begin{cases}
 \frac12\bigl(Q+(-1)^bQM\bigr),&[Q,M]=0,\\
 0,&\{Q,M\}=0,
 \end{cases}
\]
up to the harmless Pauli sign in the commuting product.  Thus branch coordinates are dyadic-rational linear combinations of previous coordinates.  Partial trace selects the coordinates with identity on the discarded subsystem.  Induction over a branch proves the ring statement.  Normalization divides all coordinates by $p_s$, itself a branch coordinate in $A(\mathcal R)$, and therefore stays in the fraction field.
\end{proof}

The same statement applies to exact channels.

\begin{corollary}[Channel field obstruction]\label{cor:channel}
Let an adaptive stabilizer protocol with resources $\mathcal R$ implement a deterministic channel $\mathcal E$.  Every entry of the Pauli transfer matrix of $\mathcal E$ lies in $A(\mathcal R)$.  For a heralded exact implementation whose nonzero success probability is independent of the input, the normalized success channel has Pauli transfer entries in $K(\mathcal R)$.
\end{corollary}

\begin{proof}
Apply the protocol to one half of the normalized maximally entangled stabilizer state $|\Phi\rangle$.  By \cref{thm:closure}, the Choi state $J(\mathcal E)=(\mathcal E\otimes I)(|\Phi\rangle\langle\Phi|)$ has Pauli coordinates in the stated ring or field.  With $d=2^n$,
\[
 2^{-n}\Tr\!\left[P\,\mathcal E(Q)\right]
 =\Tr\!\left[(P\otimes Q^{\mathsf T})J(\mathcal E)\right],
\]
so each Pauli-transfer entry is itself one of these Pauli coordinates up to the conventional Pauli transpose sign.
\end{proof}

This result is complementary to scalar resource monotones such as stabilizer extent, nullity and the dyadic monotone of \cite{Beverland2020}.  It does not count copies.  Instead, it states that a finite stabilizer protocol cannot adjoin a number absent from the field generated by its supplied resources.

\section{The dyadic Clifford-cyclotomic tower}\label{sec:tower}

Let $\C$ be the single-qubit Clifford group.  Following \cite{Forest2015}, define
\begin{equation}\label{eq:GD}
 \G_D:=\left\langle\C,U_z\!\left(\frac{\pi}{2^D}\right)\right\rangle,
 \qquad D\ge1.
\end{equation}
Since $U_z(\pi/2)$ is Clifford, $\G_1=\C$.  Modulo global phase, the Bloch representation maps these groups into $SO(3)$; we write $\widehat\G_D$ for the image.  The projective groups form a strict tower
\[
 \widehat\G_1\subsetneq\widehat\G_2\subsetneq\widehat\G_3\subsetneq\cdots,
\]
with strictness following, for example, from the fixed-axis rotation characterization in the canonical synthesis theory.

The associated real cyclotomic fields are
\begin{equation}\label{eq:KD}
 K_D:=\Q(\beta_D),
 \qquad
 \beta_D:=2\cos\frac{\pi}{2^D},
 \qquad D\ge2,
\end{equation}
with $K_1:=\Q$.  They form a quadratic tower
\begin{equation}\label{eq:tower}
 K_1\subsetneq K_2\subsetneq K_3\subsetneq\cdots,
 \qquad [K_D:K_{D-1}]=2,
\end{equation}
with
\begin{equation}\label{eq:beta}
 \beta_D^2=\beta_{D-1}+2.
\end{equation}
The ring of integers of $K_D$ is $\Z[\beta_D]$; this is the standard maximal-real-subfield description for power-of-two cyclotomic fields \cite{Washington1997}.

\begin{figure}[t]
\centering
\begin{tikzpicture}[x=2.45cm,y=0.75cm, every node/.style={font=\small},
 arr/.style={-{Latex[length=2.3mm]},thick,draw=figblue}]
\foreach \i/\lab in {1/1,2/2,3/3,4/4,5/5}{
  \node (K\i) at (\i,1) {$K_{\lab}$};
  \node (G\i) at (\i,0) {$\widehat{\G}_{\lab}$};
}
\foreach \i in {1,2,3,4}{
 \pgfmathtruncatemacro{\j}{\i+1}
 \draw[arr] (K\i)--(K\j);
 \draw[arr] (G\i)--(G\j);
}
\foreach \i in {1,2,3,4,5}{\draw[arr] (G\i)--(K\i) node[midway,right,font=\scriptsize]{entries};}
\node[align=center,anchor=west] at (5.45,0.5) {relative descent:\\$\widehat\G_D\cap SO(3,K_d)=\widehat\G_d$};
\end{tikzpicture}
\caption{The nested gate and field towers.  The relative intersection theorem proved below says that, inside the dyadic synthesizable union, descending the arithmetic field is exactly the same as descending the gate alphabet.}
\label{fig:tower}
\end{figure}
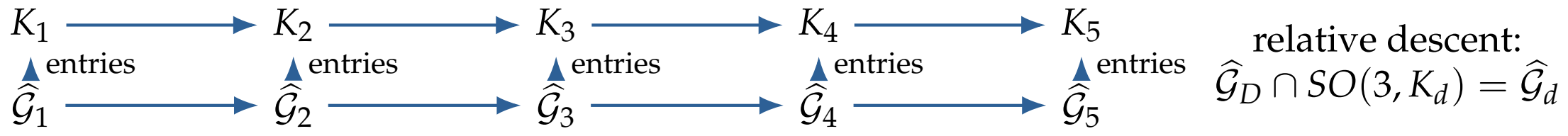

\subsection{Imported fixed-alphabet structure}
For $n=2^D$, Forest et al. give a unique canonical form for every $U\in\G_D$ and an optimal synthesis algorithm \cite{Forest2015}.  In the Bloch representation a non-Clifford element has a canonical decomposition
\begin{equation}\label{eq:forestword}
 \widehat U
 =R_{p_1}^{a_1}R_{p_2}^{a_2}\cdots R_{p_m}^{a_m}\widehat C,
\end{equation}
where $p_i\in\{x,y,z\}$, adjacent axes differ, $1\le a_i<2^{D-1}$, and $\widehat C$ is Clifford.  Let
\begin{equation}\label{eq:qa}
 q_a=2^{D-1}-2^t
 \quad\text{when }a=2^tu\text{ with }u\text{ odd}.
\end{equation}
This is the specialization of Eq.~(23) in \cite{Forest2015} to $n=2^D$.

If $N=\sum_iq_{a_i}$, their denominator-exponent theorem states that the largest $\beta_D$-denominator exponent appearing in $\widehat U$ is $N$, while the unique exceptional row has largest exponent $N-q_{a_1}$.  Their Corollary~4.2 proves uniqueness of the canonical form and minimum primitive-rotation cost at the fixed alphabet.  We use these results as imported ingredients; the new step is relative descent between consecutive dyadic alphabets.

\section{Cyclotomic parity and relative descent}\label{sec:descent}

The key arithmetic observation is that an element of the shallower field cannot have an odd reduced denominator exponent in the deeper generator.  We use the standard facts that $\mathcal O_{K_D}=\Z[\beta_D]$, that $K_D/K_{D-1}$ is quadratic, and that its nontrivial automorphism sends $\beta_D\mapsto-\beta_D$.

\begin{lemma}[Even-denominator lemma]\label{lem:even}
Let $D\ge2$.  Suppose $x\in K_{D-1}$ can be written
\[
 x=\frac{w}{\beta_D^r},
 \qquad w\in\Z[\beta_D],
\]
where $w$ is not divisible by $\beta_D$.  Then $r$ is even.
\end{lemma}

\begin{proof}
The nontrivial element $\sigma_D\in\operatorname{Gal}(K_D/K_{D-1})$ satisfies $\sigma_D(\beta_D)=-\beta_D$.  Since $x$ is fixed by $\sigma_D$,
\[
 \sigma_D(w)=(-1)^r w.
\]
If $r$ were odd, then $\sigma_D(w)=-w$.  Let $s_D=[K_D:\Q]$ and write the unique integral power-basis expression
\[
 w=c_0+c_1\beta_D+\cdots+c_{s_D-1}\beta_D^{s_D-1},
 \qquad c_j\in\Z.
\]
Applying $\sigma_D$ and comparing coefficients shows that all even-power coefficients vanish.  Hence $w=\beta_Dv$ for some $v\in\Z[\beta_D]$, contradicting the reduced-denominator assumption.
\end{proof}

\begin{remark}
The lemma also has a valuation proof.  The unique prime above $2$ is totally ramified along the power-of-two real cyclotomic tower; the valuation of an element from $K_{D-1}$ doubles upon extension to $K_D$.  The Galois proof above is shorter and keeps the argument self-contained once the integral power basis is fixed.
\end{remark}

\begin{theorem}[One-step relative intersection]\label{thm:intersection1}
For every $D\ge2$,
\[
 \boxed{\widehat\G_D\cap SO(3,K_{D-1})=\widehat\G_{D-1}.}
\]
\end{theorem}

\begin{proof}
The inclusion $\widehat\G_{D-1}\subseteq\widehat\G_D\cap SO(3,K_{D-1})$ is immediate.  For the reverse inclusion, let $\widehat U\in\widehat\G_D$ have all entries in $K_{D-1}$ and take its Forest canonical form \eqref{eq:forestword}.  If $U$ is Clifford there is nothing to prove.  Otherwise let $N=\sum_iq_{a_i}$.  By the denominator-exponent theorem of \cite{Forest2015}, some entry has reduced $\beta_D$-denominator exponent $N$, and an entry in the exceptional row has reduced exponent $N-q_{a_1}$.  Both entries lie in $K_{D-1}$, so \cref{lem:even} forces both exponents to be even.  Therefore $q_{a_1}$ is even.

Write $a_1=2^tu$ with $u$ odd.  By \eqref{eq:qa}, $q_{a_1}=2^{D-1}-2^t$.  Since $1\le a_1<2^{D-1}$, $t\le D-2$, and $q_{a_1}$ is even if and only if $t\ge1$.  Thus $a_1$ is even.  The first syllable is therefore already a shallower rotation:
\[
 R_{p_1}\!\left(\frac{a_1\pi}{2^D}\right)
 =R_{p_1}\!\left(\frac{(a_1/2)\pi}{2^{D-1}}\right)
 \in\widehat\G_{D-1}.
\]
Multiply its inverse into $\widehat U$.  Because the removed factor itself has entries in $K_{D-1}$, the remaining suffix still has entries in $K_{D-1}$; because the original Forest word was canonical, that suffix is again a canonical word at level $D$.  Induction on the number of non-Clifford syllables therefore forces every $a_i$ to be even.  Hence every non-Clifford syllable belongs to $\widehat\G_{D-1}$, as does the final Clifford, so $\widehat U\in\widehat\G_{D-1}$.
\end{proof}

\begin{corollary}[Full relative intersection]\label{cor:intersection}
For $1\le d\le D$,
\[
 \boxed{\widehat\G_D\cap SO(3,K_d)=\widehat\G_d.}
\]
\end{corollary}

\begin{proof}
Iterate \cref{thm:intersection1} from $D$ down to $d$.
\end{proof}

The statement is deliberately relative.  Forest et al. show that for most even $n$, the Clifford-cyclotomic group is strictly smaller than the ambient group of all ring-valued unitaries \cite{Forest2015}.  We do not assert that field membership alone characterizes arbitrary $SO(3,K_D)$ matrices.  Rather, \cref{cor:intersection} states that \emph{within the already-synthesizable dyadic union}, a fall in field depth is exactly a fall in gate-alphabet depth.

\section{Field depth, alphabet depth and resource depth}\label{sec:main}

We can now state the principal theorem precisely.

\begin{definition}[Field and alphabet depth]
For a projective unitary with $\widehat U\in\bigcup_D\widehat\G_D$, define
\begin{align}
 \FieldDepth(U)&:=\min\{D\ge1:\widehat U\in SO(3,K_D)\},\\
 \AlphabetDepth(U)&:=\min\{D\ge1:\widehat U\in\widehat\G_D\}.
\end{align}
Both quantities depend only on the channel $\rho\mapsto U\rho U^\dagger$.
\end{definition}

\begin{theorem}[Arithmetic alphabet theorem]\label{thm:FA}
For every $\widehat U\in\bigcup_D\widehat\G_D$,
\[
 \boxed{\FieldDepth(U)=\AlphabetDepth(U).}
\]
\end{theorem}

\begin{proof}
Every generator of $\widehat\G_D$ has entries in $K_D$, so $\FieldDepth(U)\le\AlphabetDepth(U)$.  Let $D=\AlphabetDepth(U)$ and $d=\FieldDepth(U)$.  If $d<D$, then $\widehat U\in\widehat\G_D\cap SO(3,K_d)$.  By \cref{cor:intersection}, $\widehat U\in\widehat\G_d$, contradicting the minimality of the projective alphabet depth $D$.
\end{proof}

\subsection{Dyadic program-state resource depth}
For $D\ge2$, let $\mathcal M_D$ denote the set of one-qubit dyadic program states $q_{m,D}$ of \eqref{eq:qmd} with $m$ odd, together with their Clifford images.  Their Pauli coordinates generate exactly $K_D$: indeed
\[
 \langle X\rangle=\sin\frac{m\pi}{2^D},
 \qquad
 \langle Z\rangle=\cos\frac{m\pi}{2^D},
\]
and for odd $m$ these values are Galois conjugates of the primitive real cyclotomic generators.

\begin{definition}[Dyadic program-state resource depth]
For a non-Clifford channel with $\widehat U\in\bigcup_D\widehat\G_D$, let $\ResourceDepth(U)$ be the least $D\ge2$ for which there exists a finite adaptive stabilizer protocol that implements $\rho\mapsto U\rho U^\dagger$ exactly using an arbitrary finite number of resource states from $\bigcup_{d\le D}\mathcal M_d$.  Stabilizer ancillas, adaptive Pauli measurement, classical feed-forward, discarding, and catalytic use of states from the same declared resource family are allowed.
\end{definition}

\begin{theorem}[Exact resource-depth theorem]\label{thm:triple}
For every non-Clifford channel with $\widehat U\in\bigcup_D\widehat\G_D$,
\[
 \boxed{\FieldDepth(U)=\AlphabetDepth(U)=\ResourceDepth(U).}
\]
\end{theorem}

\begin{proof}
Set $D_*=\FieldDepth(U)=\AlphabetDepth(U)$ by \cref{thm:FA}.  For the lower bound, suppose $U$ had an exact stabilizer implementation whose every dyadic resource state had depth at most $d<D_*$.  All resource Pauli coordinates would then lie in $K_d$, and because the fields are nested, the field generated by any finite collection of them is still contained in $K_d$.  By \cref{cor:channel}, every Pauli-transfer entry of the implemented channel would lie in $K_d$.  The Pauli transfer matrix of a unitary channel is its Bloch rotation, contradicting the definition of $D_*$.  Thus $\ResourceDepth(U)\ge D_*$.

For the upper bound, $\widehat U\in\widehat\G_{D_*}$ has a finite exact decomposition, up to global phase, into Clifford gates and primitive $U_z(\pi/2^{D_*})$ rotations.  By \cref{lem:SH} each primitive rotation has a Clifford-equivalent dyadic program state of depth $D_*$.  The injection rule \eqref{eq:inj0}--\eqref{eq:inj1} implements the rotation exactly; every failed branch requests only a doubled-angle correction, hence a strictly shallower dyadic layer until the correction is Clifford.  Therefore depth $D_*$ resources suffice, and $\ResourceDepth(U)\le D_*$.
\end{proof}

This theorem should not be confused with a statement that resource \emph{count} equals arithmetic depth.  The count is a separate optimization problem.  The theorem identifies the deepest algebraic layer that must appear somewhere in an exact implementation.

\section{Exact resource effects and optimal compilation}\label{sec:type}

At a fixed alphabet $n=2^D$, Forest et al. define $\Tcost_n(U)$ as the minimum number of primitive $U_z(\pi/n)$ gates required to implement $U$ up to global phase and prove that their canonical decomposition attains this minimum \cite{Forest2015}.  Combining their fixed-level theorem with \cref{thm:triple} yields a two-component exact resource signature.  We call it a resource effect rather than a full type system: the present paper does not supply general typing rules for an arbitrary higher-order quantum language.

\begin{definition}[Exact dyadic resource effect]
For non-Clifford $U\in\bigcup_D\G_D$, let
\[
 D_*(U):=\FieldDepth(U)=\AlphabetDepth(U)=\ResourceDepth(U)
\]
and define
\[
 \operatorname{QEff}(U)
 :=\left(D_*(U),\Tcost_{2^{D_*(U)}}(U)\right).
\]
\end{definition}

\begin{theorem}[Lexicographic optimality]\label{thm:lex}
Order exact dyadic gate implementations first by alphabet depth $D$ and then by primitive-rotation count $T$.  The pair $\operatorname{QEff}(U)=(D_*,T_*)$ is the unique lexicographic optimum: no exact implementation exists with depth $D<D_*$, and among depth-$D_*$ implementations none uses fewer than $T_*$ primitive rotations.
\end{theorem}

\begin{proof}
The first claim is the definition of $\AlphabetDepth$ together with \cref{thm:triple}.  The second is the fixed-alphabet optimality theorem of \cite{Forest2015} applied at $n=2^{D_*}$.
\end{proof}

The result gives a static effect analysis familiar from resource-aware programming languages, but with a stronger guarantee than an upper bound: the first component is an exact operational minimum, not merely a conservative estimate.  This complements quantum type systems that track stabilizer structure, $T$ gates and injection patterns \cite{Sundaram2022}, and resource type-and-effect systems that infer quantitative circuit bounds \cite{Colledan2025}.

\subsection{Exact field-depth analysis}
Suppose a source circuit gives an a priori alphabet bound $D_{\max}$, so its exact Bloch matrix lies in $SO(3,K_{D_{\max}})$.  Field depth can then be computed by fixed-field descent rather than by searching gate alphabets.  Starting from $j=D_{\max}$, apply the nontrivial involution $\sigma_j:K_j\to K_j$ with $\sigma_j(\beta_j)=-\beta_j$ entrywise.  If every Bloch entry is fixed, then the matrix lies in $SO(3,K_{j-1})$ and one decrements $j$; otherwise the current $j$ is the field depth.  Because the source supplies a finite upper bound and exact algebraic representations, this process terminates.  By \cref{thm:FA,thm:triple}, its output is simultaneously the minimum projective gate alphabet and the minimum dyadic program-state layer.

\subsection{Compilation pipeline}
An exact compiler for the dyadic single-qubit language can therefore be organized as follows.
\begin{enumerate}[label=\textbf{Stage \arabic*.}]
\item Compute the Bloch matrix $\widehat U$ exactly from the source program.
\item Starting from the source-level upper bound, descend the cyclotomic tower by the fixed-field test above.  The first non-fixed level is the exact field depth; by \cref{thm:FA,thm:triple}, it is also the minimum projective gate alphabet and minimum dyadic program-state resource depth.
\item Run the fixed-alphabet canonical synthesis of \cite{Forest2015} once at $n=2^D$ to obtain minimum primitive-rotation count.
\item Replace primitive rotations, if a measurement-based backend is requested, by the depth-$D$ program states of \cref{sec:mist} and their adaptive correction cascades.
\item Verify the emitted program by exact algebraic channel semantics, for example through an exact Choi representation.
\end{enumerate}

No search through all shallower gate alphabets is required once field depth has been identified.

\section{Faithful axial embedding and a universal separating test}\label{sec:fullabs}

The main theorem concerns the entire dyadic Clifford-cyclotomic union.  The original mist syntax gives a particularly simple source-to-target semantics on the axial fragment.  Let source terms be normalized-midpoint expressions with denotation $x\in\Z[1/2]\cap[0,1]$, and compile
\[
 \Compile(x):=\mathcal U_x,
 \qquad
 \mathcal U_x(\rho)=U_z(\pi x)\rho U_z(\pi x)^\dagger.
\]
By \cref{lem:SH}, the source state $q(x)$ is a free-Clifford transform of the phase-program state that injects $U_z(\pi x)$.

\begin{theorem}[Semantic faithfulness on the axial fragment]\label{thm:fullabs}
For dyadic $x,y\in[0,1]$,
\[
 q(x)=q(y)
 \quad\Longleftrightarrow\quad
 \mathcal U_x=\mathcal U_y.
\]
Moreover, one fixed stabilizer observation separates every inequivalent pair: prepare $|+\rangle$, apply the compiled channel, and measure $X$.
\end{theorem}

\begin{proof}
The parametrization $q(x)$ is injective on $[0,1]$, and $U_z(\pi x)$ and $U_z(\pi y)$ define the same channel on that interval only when $x=y$.  For the separating experiment, the probability of the $+1$ outcome after applying $U_z(\pi x)$ to $|+\rangle$ is
\[
 p_+(x)=\frac{1+\cos(\pi x)}{2},
\]
which is strictly decreasing on $[0,1]$.
\end{proof}

This theorem is intentionally stated as faithfulness rather than as a general full-abstraction result: the midpoint source fragment used here has no independently developed contextual operational semantics.  Fully abstract quantum models are established prior art \cite{SelingerValiron2008}; the present point is the exact, experimentally separable embedding of this particular source notation into channel semantics.

\section{Examples and structural consequences}\label{sec:examples}

\subsection{The Hadamard magic state}
For $x=1/4$, \cref{thm:dyadicdepth} gives source depth $D=2$ and
\[
 q_{1,2}=\cos\frac\pi8|0\rangle+\sin\frac\pi8|1\rangle.
\]
The corresponding phase gate is $U_z(\pi/4)$, i.e. the $T$ gate up to the conventional choice of global phase.  Thus the nested Hadamard eigenstate from \cite{German2026,NewtonGerman2026} is the first primitive non-Clifford dyadic program state in this hierarchy.

\subsection{Unlimited shallower states cannot cross an exact field boundary}
Let $D\ge3$.  Any finite collection of program states of depths at most $D-1$ has Pauli coordinates in $K_{D-1}$.  By \cref{thm:closure}, an adaptive stabilizer protocol using any finite number of such states still produces only channels with Pauli-transfer entries in $K_{D-1}$.  Therefore it cannot exactly implement a target $U$ with $\FieldDepth(U)=D$.  This is an exact-arithmetic obstruction, not a statement about approximate synthesis: approximation can of course approach finer rotations using a fixed universal alphabet.

\subsection{Resource depth survives Clifford-hierarchy escape}
The Clifford hierarchy is not closed under multiplication above its second level.  For example, products of noncommuting dyadic axial rotations may leave every finite hierarchy level even though they remain exactly synthesizable in a finite Clifford-cyclotomic alphabet.  The complete one-qubit hierarchy classification in \cite{deSilvaLautsch2025} makes the distinction explicit.  Our type $D_*(U)$ remains finite because it is defined by exact arithmetic/gate-alphabet reachability rather than finite-hierarchy membership.  Thus ``Clifford level'' and ``resource depth'' agree on primitive axial rotations but are different invariants for compositional programs.

\section{Scope, claim boundary and related work}\label{sec:scope}

The paper deliberately separates imported results from new deductions.

\paragraph{Imported ingredients.}
The mist-state language and its fixed-point examples originate in \cite{German2026}; the certified semantics, explicit normalization discipline, CANF and dyadic-geodesic closure are from \cite{NewtonGerman2026}.  Magic-state injection and small-angle rotation resources are established technology \cite{CampbellOGorman2016}; the recent general magic-gate-teleportation framework of Zheng, Zang and Kubica gives a broad structural account of useful resource states and feed-forward \cite{ZhengZangKubica2026}.  Forest et al. supply the fixed-alphabet canonical form, denominator-exponent pattern and rotation-count optimality \cite{Forest2015}, building on earlier structure theory for discrete rotation groups such as Radin--Sadun \cite{RadinSadun1998}.  Arithmetic-group work studies the ambient Clifford-cyclotomic groups and their indices \cite{Ingalls2019}, while recent multiqubit work characterizes power-of-two ring synthesis with ancillas \cite{Amy2024,DinhRoss2025}.  Number-theoretic resource lower bounds for stabilizer computation, including a dyadic monotone, are established in \cite{Beverland2020}.  Dyadic rewriting/completeness is also known in sum-over-paths form \cite{Vilmart2024}.  None of those results is re-claimed here.

\paragraph{Contribution isolated here.}
The mathematical step specific to this paper is the relative descent theorem \cref{thm:intersection1,cor:intersection}, obtained by combining the fixed-alphabet denominator pattern with the even-denominator lemma for the power-of-two real cyclotomic tower.  Its principal consequence is the equality between field depth and minimum gate-alphabet depth, \cref{thm:FA}.  Combining that equality with the stabilizer arithmetic-closure theorem yields the program-state resource equality \cref{thm:triple}.  Finally, fixed-alphabet optimality converts the common depth into the lexicographically optimal resource effect of \cref{thm:lex}.

\paragraph{Limitations.}
The resource equality is an exact, single-qubit, dyadic statement in a declared stabilizer-plus-dyadic-program-state model.  It does not claim that field depth measures the amount of magic, that it determines approximate synthesis cost, or that the same equality holds for arbitrary multiqubit magic resources.  It also does not replace existing complete rewriting calculi or general type systems, and $\operatorname{QEff}$ should not be read as a principal type theorem for a higher-order language.  The result instead identifies an exact arithmetic layer that controls reachability and a certifiable static resource effect in one nested single-qubit resource family.

\section{Conclusion}

A simple dyadic source syntax leads to a stronger invariant than its educational origin suggests.  Normalized midpoint depth identifies a canonical hierarchy of one-qubit program states; those states inject exact dyadic phase rotations; and their Pauli expectations generate the same real cyclotomic tower that appears in Clifford-cyclotomic synthesis.  The stabilizer arithmetic-closure theorem turns this field tower into an operational obstruction.  The relative intersection theorem then supplies the missing converse: within the exactly synthesizable dyadic union, descending the arithmetic field is precisely descending the gate alphabet.

The resulting identity
\[
 \FieldDepth(U)=\AlphabetDepth(U)=\ResourceDepth(U)
\]
provides an exact resource effect for non-Clifford single-qubit dyadic programs in the declared program-state model.  It remains meaningful after composition leaves the finite Clifford hierarchy, and when paired with fixed-alphabet optimal synthesis it yields a lexicographically optimal compiler objective.  The next natural question is whether an analogous relative-descent theorem survives in selected multiqubit Clifford-cyclotomic or commuting-Pauli fragments, where field composita, stabilizer codes and nontrivial gate-teleportation structure interact.

\appendix
\section{Exact channel normal form for adaptive branches}\label{app:choi}

For completeness we record a representation suitable for an independent verifier.  Let a finite adaptive program have externally visible classical outcome $o$ and internal measurement history $s$, with branch Kraus operator $K_{o,s}$.  Define
\[
 J_o:=\sum_s |\vecop K_{o,s}\rangle\langle\vecop K_{o,s}|.
\]
The family $\{o\mapsto J_o\}$ is the Choi representation of the externally visible quantum instrument after internal transcripts have been hidden.  Equality of these Choi operators is equivalent to equality of the induced completely positive maps.  In the dyadic fragment all entries lie in an explicitly generated cyclotomic number field, so equivalence is an exact algebraic decision rather than a floating-point comparison.

For the injection circuit of \cref{sec:mist}, each terminal branch is proportional to the same target unitary after the prescribed feed-forward.  Summing branch Choi matrices therefore gives the Choi matrix of the target unitary channel exactly.  This supplies a convenient backend certificate for the compiler pipeline of \cref{sec:type}.

\section{A second proof of the parity lemma}\label{app:valuation}

The Galois proof of \cref{lem:even} can be replaced by a valuation argument.  Let $\mathfrak p_D$ be the unique prime of $K_D$ over $2$.  In the power-of-two cyclotomic tower the extension $K_D/K_{D-1}$ is quadratic and $\mathfrak p_{D-1}$ is totally ramified, so for $x\in K_{D-1}$,
\[
 v_{\mathfrak p_D}(x)=2v_{\mathfrak p_{D-1}}(x).
\]
The principal ideal generated by $\beta_D$ is supported at $\mathfrak p_D$; hence the reduced $\beta_D$-denominator exponent of an element from the base field is even.  This proves \cref{lem:even} independently of coefficient parity.  Standard facts about cyclotomic ramification may be found in \cite{Washington1997}; related denominator arithmetic is developed in the appendices of \cite{Forest2015}.

\section{Dependency map}\label{app:dependency}

The proof dependencies are deliberately short:
\[
\begin{array}{c}
\text{dyadic midpoint theorem}\Rightarrow\text{program-state depth}\Rightarrow\text{injection upper bound},\\[1mm]
\text{stabilizer arithmetic closure}\Rightarrow\text{field lower bound},\\[1mm]
\text{Forest denominator pattern}+\text{even-denominator lemma}\Rightarrow\text{relative intersection},\\[1mm]
\text{relative intersection}\Rightarrow\FieldDepth=\AlphabetDepth,\\[1mm]
\text{field lower bound}+\text{injection upper bound}\Rightarrow\ResourceDepth=\FieldDepth,\\[1mm]
\text{fixed-alphabet optimality}\Rightarrow\text{lexicographically optimal resource effect}.
\end{array}
\]
This organization also marks the prior-art boundary: only the implications stated in the present paper depend on the new relative-descent argument; the fixed-alphabet synthesis theorem itself is imported from \cite{Forest2015}.


\begin{thebibliography}{99}

\bibitem{German2026}
D.-A. German,
\newblock A term-rewriting semantics for pure quantum states,
\newblock \emph{arXiv:2607.06584}, 2026.

\bibitem{NewtonGerman2026}
T. Newton and D.-A. German,
\newblock Certified misty-state rewriting: A question-and-answer guide,
\newblock \emph{arXiv:2608.10024}, 2026.

\bibitem{Forest2015}
S. Forest, D. Gosset, V. Kliuchnikov, and D. McKinnon,
\newblock Exact synthesis of single-qubit unitaries over Clifford-cyclotomic gate sets,
\newblock \emph{Journal of Mathematical Physics} 56, 082201 (2015),
\newblock doi:10.1063/1.4927100; arXiv:1501.04944.

\bibitem{CampbellOGorman2016}
E. T. Campbell and J. O'Gorman,
\newblock An efficient magic state approach to small angle rotations,
\newblock \emph{Quantum Science and Technology} 1, 015007 (2016),
\newblock arXiv:1603.04230.

\bibitem{ZhengZangKubica2026}
Y. Zheng, A. Zang, and A. Kubica,
\newblock Magic gate teleportation: Structure, useful resource states, and simpler feedforward,
\newblock \emph{arXiv:2607.08508}, 2026.

\bibitem{Beverland2020}
M. Beverland, E. Campbell, M. Howard, and V. Kliuchnikov,
\newblock Lower bounds on the non-Clifford resources for quantum computations,
\newblock \emph{Quantum Science and Technology} 5, 035009 (2020),
\newblock arXiv:1904.01124.

\bibitem{deSilvaLautsch2025}
N. de Silva and O. Lautsch,
\newblock The Clifford hierarchy for one qubit or qudit,
\newblock \emph{Proceedings of the Royal Society A} 481, 20250035 (2025),
\newblock arXiv:2501.07939.

\bibitem{Vilmart2024}
R. Vilmart,
\newblock Rewriting and completeness of sum-over-paths in dyadic fragments of quantum computing,
\newblock \emph{Logical Methods in Computer Science} 20(1):20 (2024),
\newblock arXiv:2307.14223.

\bibitem{Sundaram2022}
A. Sundaram, R. Rand, K. Singhal, and B. Lackey,
\newblock A rich type system for quantum programs,
\newblock \emph{arXiv:2101.08939}, version 3, 2022.

\bibitem{Colledan2025}
A. Colledan, U. Dal Lago, and N. Vazou,
\newblock Circuit width estimation via effect typing and linear dependency,
\newblock \emph{ACM Transactions on Programming Languages and Systems} 47(3), 2025,
\newblock doi:10.1145/3737282.

\bibitem{SelingerValiron2008}
P. Selinger and B. Valiron,
\newblock On a fully abstract model for a quantum linear functional language,
\newblock \emph{Electronic Notes in Theoretical Computer Science} 210, 123--137 (2008),
\newblock doi:10.1016/j.entcs.2008.04.022.


\bibitem{RadinSadun1998}
C. Radin and L. Sadun,
\newblock Subgroups of $SO(3)$ associated with tilings,
\newblock \emph{Journal of Algebra} 202, 611--633 (1998),
\newblock arXiv:math/9601202.

\bibitem{Ingalls2019}
C. J. Ingalls, B. W. Jordan, A. Keeton, A. Logan, and Y. Zaytman,
\newblock The Clifford-cyclotomic group and Euler--Poincar\'e characteristics,
\newblock \emph{arXiv:1903.09497}, 2019.

\bibitem{Amy2024}
M. Amy, A. N. Glaudell, S. Kelso, W. Maxwell, S. S. Mendelson, and N. J. Ross,
\newblock Exact synthesis of multiqubit Clifford-cyclotomic circuits,
\newblock \emph{arXiv:2311.07741}, revised 2024.

\bibitem{DinhRoss2025}
L. Dinh and N. J. Ross,
\newblock Contributions to the theory of Clifford-cyclotomic circuits,
\newblock \emph{Electronic Proceedings in Theoretical Computer Science} 426, 231--246 (2025),
\newblock arXiv:2508.14674.

\bibitem{Washington1997}
L. C. Washington,
\newblock \emph{Introduction to Cyclotomic Fields}, 2nd ed.,
\newblock Springer, 1997.

\end{thebibliography}
\end{document}